# Temperature-Driven Topological Transition in 1T'-MoTe$_2$

Ayelet Notis Berger[1,2]\*, Erick Andrade[1]\*, Alex Kerelsky[1], Drew Edelberg[1], Jian Li[3], Zhijun Wang[4], Lunyong Zhang[5,6], Jaewook Kim[6], Nader Zaki[7], Jose Avila[8], Chaoyu Chen[8], Maria C Asensio[8], Sang-Wook Cheong[6], Bogdan A. Bernevig[4,9,10,11], Abhay N. Pasupathy[1]

[1] Department of Physics, Columbia University, New York, NY 10027. [2] Department of Materials Science, Columbia University, New York, NY 10027. [3] Westlake Institute for Advanced Study, Hanzhou, China. [4] Department of Physics, Princeton University, Princeton, NJ 08520. [5] Laboratory for Pohang Emergent Materials & Max Plank POSTECH Center for Complex Phase Materials, Max Planck POSTECH/Korea Research Initiative , Pohang 790-784, Korea. [6] Rutgers Center for Emergent Materials & Department of Physics and Astronomy, Rutgers University, Piscataway, NJ 08854. [7] Department of Applied Physics & Applied Mathematics, Columbia University, New York, NY 10027. [8] Synchrotron SOLEIL Orme des Merisiers – Saint Aubin BP 48 – 91192 – GIF SUR YVETTE Cedex FRANCE [9] Laboratoire Pierre Aigrain, Ecole Normale Supérieure-PSL Research University, CNRS, Université Pierre et Marie Curie-Sorbonne Universités, Université Paris Diderot-Sorbonne Parice Cité, 24 rue Lhomond, 75231 Paris Cedex 05, France† [10] Donostia International Physics Center, P. Manuel de Lardizabal 4, 20018 Donostia-San Sebastián, Spain.†

[11] Sorbonne Universités, UPMC Univ Paris 06, UMR 7589, LPTHE, F-75005, Paris, France. †

\* These authors contributed equally to this work.

†On sabbatical

**Abstract:**

The topology of Weyl semimetals requires the existence of unique surface states. Surface states have been visualized in spectroscopy measurements, but their connection to the topological character of the material remains largely unexplored. 1T'-MoTe$_2$, presents a unique opportunity to study this connection. This material undergoes a phase transition at 240K that changes the structure from orthorhombic (putative Weyl semimetal) to monoclinic (trivial metal), while largely maintaining its bulk electronic structure. Here we show from temperature-dependent quasiparticle interference measurements that this structural transition also acts as a topological switch for surface states in 1T'-MoTe$_2$. At low temperature, we observe strong quasiparticle scattering, consistent with theoretical predictions and photoemission measurements for the surface states in this material. In contrast, measurements performed at room temperature show the complete absence of the scattering wavevectors associated with the trivial surface states. These distinct quasiparticle scattering behaviors show that 1T'-MoTe$_2$ is ideal for separating topological and trivial electronic phenomena via temperature dependent measurements.

**Introduction:**

The Weyl fermion [1] is a massless chiral solution of the Dirac equation. Several solid-state materials whose crystal structures break time reversal or inversion symmetry have been predicted [2-11] and observed [12-40] to host quasiparticle excitations that mimic free Weyl fermions. In these Weyl semimetals, Weyl points exist as a touching between a hole and an electron pocket. While fundamental Weyl fermions are constrained by the Dirac equation to move at the speed of light, in the solid-state context the relationship between energy and momentum of the quasiparticles is determined by the material-specific bandstructure. In a type I Weyl semimetal, the hole and electron pockets which touch to form the Weyl points do not overlap in energy except at the Weyl points, leading to a bandstructure where quasiparticles exactly mimic fundamental Weyl fermions. In a type II Weyl semimetal however, the hole and electron pockets, which form the Weyl points, overlap over a range of energies. Type II Weyl fermions are only possible as a quasiparticle excitation in a system where Lorentz invariance is broken [6], and thus are uniquely different from free Weyl fermions. Much recent activity has centered around the finding that at the surface of a Weyl semimetal, Weyl points are necessarily connected in pairs of opposite chirality by Fermi arc surface states, which form open contours on the Fermi surface [5,41]. Accordingly, most spectroscopic experiments have concentrated on observing the surface states in these materials. The majority of these experiments have focused on type I Weyl semimetals [13-16,18-27,34,42], with a smaller number of observations on type II Weyl semimetals including $MoTe_2$ [30-32,35,36,40] and its alloys with $WTe_2$ [33,37-39,43]. In spectroscopy measurements it is possible to conclusively prove the surface nature of electronic states; however, proving that the surface states are topologically protected arcs is a much more delicate task. An alternative way of studying Fermi surface properties is via transport. Many unique transport characteristics have been predicted for the Weyl band-structures in general and the Fermi arc states in particular [44-51]. Transport experiments on Weyl semimetals have indeed seen several intriguing features [17,52-60]; however, the link to the Weyl band-structure is often difficult to make. To clarify this connection, it would be ideal to use a system where Weyl behavior can be switched on and off using an experimentally tunable parameter like temperature, pressure or applied field.

One recently discovered type-II Weyl semimetal is MoTe$_2$, a transition metal dichalcogenide that exists as several different polytypes at high temperature [61,62]. At room temperature, the crystal structure is either hexagonal (2H, or α phase) or monoclinic (1T', or β phase). When the monoclinic phase is cooled it undergoes a structural transition at 250K to become orthorhombic, known as the Td phase [61,62]. In the Td phase, each Mo atom is at the center of a buckled Te octahedron, as seen in Figure 1. The Td phase breaks inversion symmetry, and theoretical predictions of the electronic structure of this phase indicate that it is a topological type-II Weyl semimetal, although the number of Weyl points and the location of the associated Fermi arcs are very sensitive to the exact lattice parameters [7,8]. On the other hand, in the monoclinic 1T' phase, inversion symmetry is restored and the material is topologically trivial. Thus, the cooling or warming of a sample across the orthorhombic-monoclinic phase boundary offers a clean and simple way to compare the electronic properties in the topologically trivial and non-trivial phases. In this work, we demonstrate this switching behavior by visualizing changes in electronic structure across the temperature boundary. We supplement our experimental measurements with theoretical calculations and ARPES measurements of the band structure.

**Results:**

Our STM and ARPES measurements are performed on crystals grown by the flux method (details in Methods) and quenched from high temperature to preserve the metastable 1T' phase at room temperature. Measurements are performed on in-situ cleaved crystals. STM topographic measurements taken over large areas (Figure 1c) show clean, flat surfaces with several defects present at both the chalcogen and the metal sites. Figure 1d shows two-dimensional cuts of the band-structure measured by ARPES along the cuts X-Gamma-Y over a wide range of energy. Our measurements are consistent with theoretical calculations and previous measurements [30-32] on similar crystals indicating that the samples are indeed in the Td phase.

In STS imaging measurements, real-space differential conductance images can be connected to the material bandstructure in the presence of scattering from crystal defects. Point defects in materials can act as elastic scattering centers, giving rise to quasiparticle interference (QPI) patterns that are located around each impurity. The wavevectors present in the interference pattern are related to the momentum transfer provided by the defect in the scattering process. Shown in figure 1e and f are two real space images that are obtained at -35 mV and 50 mV respectively. We find that the scattering from each individual defect in Td-MoTe$_2$ is weak, but the collective effect of all the scattering centers in the material causes clear modulations in the local density of states at each energy. This situation is reminiscent of similar experiments in other topological insulator materials [63]. The wavevectors present in the real space image can be seen by taking a magnitude Fourier transform of the image, as shown in the inset to the real space images in Figures 1e,f (see supplementary information, section III, for more detailed spectroscopy maps). The QPI wavevectors we observe in Td-MoTe$_2$ have two distinctive features. First, as can be seen in the inset to figure 1e, we observe horizontal "wings" aligned with the a-axis of the crystal (red box, inset to figure 1e). Second, we observe a vertical stripe-like feature as seen in the inset of figure 1f (green box). This feature is aligned with the b-axis of the crystal. We describe the evolution of the intensity and wavevectors of the wings and the vertical stripes in detail below.

Our experimental results for the observed wavevectors from QPI can be compared to predictions based on theoretical calculations of the surface electronic structure. Our DFT calculations [7] using lattice parameters from x-ray diffraction at 100K show four Weyl points at k=(0.1011, 0.0503, 0) and points related by the reflections $M_{x,y}$. Each pair of these Weyl points located at the same value of $k_x$ is necessarily connected by Fermi arc surface states. In addition to these arcs, trivial surface states also exist in the same region in momentum space. We can calculate the surface bandstructure of the material at each energy either by projecting the full density-functional theory (DFT) calculation onto the surface, or from a tight-binding model that is constructed from DFT calculations (see methods). The computed surface bandstructure at three energies is shown in Figure 2b-d. A number of bulk and surface bands are seen in these calculations. Since the surface states are truly two-dimensional, they can be easily picked out from the other contributions to the surface bandstructure by a simple process of thresholding the surface spectral density, as illustrated in figure 2b-d. The color map chosen uses shades of blue, with only the most intense values colored red. The surface states contrast sharply with their surrounding bulk bands located near the same wavevectors, so this color map easily identifies the surface states based on intensity. The identification of the surface states by this procedure allows us to track the evolution of these states in energy, as shown in figure 2a. We see that the surface bands are fairly localized in energy – they first appear around -80 meV, then lose intensity and hybridize strongly with the bulk bands above -10 meV. The evolution of hole and electron pockets (highlighted with brown and green lines) matches well with that found in measurements from ARPES, as shown in figure 2e-g. Our ARPES results are consistent with previous measurements [30-32] of the same material. The Weyl points themselves are not visualized in the data since they occur above the Fermi level [7].

We note that calculations using a slightly different lattice constant (~1%) [8] produce different results for the surface state bandstructure. In our calculations [7], there is one Weyl point in each quadrant of the BZ, located at at ($\pm$0.1011a*,$\pm$0.0503b*,0). The large surface state connects the Weyl points across the $k_y$ axis and is a topological Fermi arc. In the other calculation[8], there are two Weyl points in each quadrant of the BZ at ($\pm$0.1024a*,$\pm$0.0128b*,0) and (0.1001a*,0.0530b*,0). The Fermi arcs are much smaller in extent and connect the Weyl points located within each quadrant. A large, singly degenerate surface state is seen in this calculation as well, although it is topologically trivial in this calculation. The origin of this large surface state is currently not understood theoretically – the fact that it is single degenerate is quite unusual, since its Rashba partner seems to be missing in the surface bands

[35]. While the topological properties of the surface bands are quite different in the two calculations, the actual extent and dispersion of the surface bands themselves are very similar. We will comment further on the difference below, but we will proceed by comparing STM experiments to the calculations from Wang, et al[7].

Using our theoretically calculated bandstructure, we can now calculate the expected scattering wavevectors in a QPI experiment. An elastic scattering event from a defect results in a change in momentum $q$ of a quasiparticle incident on it. Real space interference patterns and their Fourier transforms will show intensity only at values of $q$ allowed by the bandstructure at each energy [64-66]. Additional selection rules that govern the relative intensities for different $q$ can exist due to crystal and time-reversal symmetry conservation laws [63], structure of the scattering potential [66,67], or matrix elements. In our experiments, we consider the impurities giving rise to scattering processes to be largely non-magnetic, as observed for other transition-metal dichalcogenides [66]. To compute the expected QPI pattern at every energy, we calculate the spin-conserved scattering probability (SSP)

$$J_s(q) = \frac{1}{2} \sum_k \sum_{i=0,1,2,3} \rho_i(k) \rho_i(k+q) \qquad (1)$$

where $\rho_0(k)$ stands for charge densities, and $\rho_i(k)$ ($i = 1,2,3$) stands for spin densities along three orthogonal orientations [see Methods].

While the experimental QPI occurs at allowed values of $q$ from the theory, only a subset of the theoretically allowed values is actually observed in experiment. It is therefore worth understanding the sources within theory of the features that are seen in experiment. In general, the surface band-structure has a part that comes from the surface states (red band in Figure 3a), and a remainder that comes from the projections of various bulk bands. A scattering process can therefore take place between bulk bands alone, from a bulk to a surface band, or between the surface bands alone. We decompose the theoretical bandstructure (3a) into surface and bulk bands to calculate the three separate QPI components, shown in figure 3d-f. By comparing each of these separately with the experimental QPI in figure 3c, we see that the most distinctive feature of the experimental QPI (the "wings" denoted by the circled region) is seen prominently in the surface-surface scattering as well as (to a lesser extent) in the surface-bulk scattering. On the other hand, the bulk-bulk scattering does not show any particular features at this intensity, and indeed several of the features seen in the bulk scattering do not exist in the experiment. Thus, on the basis of the match between theory and experiment, we can identify the "wings" as being associated with surface state scattering.

Next, we consider the dispersion of the QPI features with energy. Shown in figure 4a-f are the expected QPI from theory, color-coded as before to distinguish the various contributions to scattering. Inspection of this energy evolution shows two distinctive features associated with the surface state scattering, coming from inter-arc and intra-arc scattering, respectively. The inter-arc scattering that has already been discussed above leads to the "wings" along the $k_x$ direction that disperse along the y direction with increasing energy, as a consequence of similar dispersion in the surface bands themselves. The second prominent feature observed in the surface state scattering is intra-arc scattering, which is seen for small momentum transfers. At the lowest energies where the arcs are observed (-80 mV), the intra-arc scattering is difficult to distinguish from strong bulk scattering that is present at the same wavevectors (figure 4m). However, as the energy increases (going from figure 4a to 4f), the intra-arc scattering disperses outwards along $k_y$, while at the same time the amount of bulk scattering present at these wavevectors diminishes. Since the inter-arc and intra-arc scatterings appear at very different wavevectors, we perform two separate imaging experiments with different real-space size scales to see the two scatterings clearly. An additional complication in the comparison between theory and experiment is the presence of Fermi level shifts in the sample as a function of spatial position or time. Such shifts, which have been documented to occur because of surface doping [30,68], sample aging [69], chemical inhomogeneity [70] or non-stoichiometry [57] in both topological insulators [68-70] and semimetals [30,57], imply that an energy shift has to be applied to the experimental data before comparison with theory. Shown in figure 4a-f and figure 4g-l are two sets of experimental data that are acquired to make comparison with the inter- and intra-arc scattering features respectively (see supplementary information, section IV, for details of energy alignment between the two data sets). Imaging performed over a larger real-space area (4g-l) reveals the intra-arc scattering wavectors. While the inter-arc scattering is also observed weakly at this real space resolution (see supplementary sections III, IV for more details), imaging performed on a smaller area (figures 4a-f) reveals the large wavevector, inter-arc scattering ("wings") more clearly. We see from these data sets that the experimental dispersion of the QPI features matches reasonably well with both the intra-arc and inter-arc scatterings.

    While our low temperature STS measurements are in good agreement with our surface state calculations, we have explained earlier that the calculations themselves are extremely sensitive to the lattice parameters used. Theoretically, the large surface state from which we see QPI can be rendered topological or not by changing the small $k_y$ parts of the band structure – however these changes are very hard to detect in QPI or other spectroscopic experiments due to limited experimental resolution. Thus, detecting the surface states at low temperature alone is not rigorous enough to claim agreement with the theoretical Fermi arcs. In our experiments, we are able to use the temperature-driven phase transition to gain crucial additional insight into the topological character of 1T'-MoTe$_2$. The orthorhombic-monoclinic structural transition involves a small (~ 3 degrees) distortion of the c-axis stacking while otherwise leaving the structure intact. As a consequence, the bulk bandstructure is only very slightly changed between the two phases. This is shown in figure 5a and 5b, which are the surface-projected bandstructures at -60 meV calculated by tight binding for both phases. We can see that the overall bulk bandstructure is very similar for both phases. However, an important consequence of the phase transition is that inversion symmetry is recovered in the monoclinic phase, implying that the high temperature phase does not have Weyl points with topologically protected Fermi arc surface states. Topologically trivial surface states can still exist in the material, and a close examination of figure 5b

interestingly shows that such a trivial surface state does exist at high temperature. The surface states at low and high temperature are highlighted on the $k_x<0$ side of figure 5a-b in brown (further details of the dispersion of the band structure in both phases is provided in the supporting information, section VIII).

We can calculate the expected QPI based on our surface projected tight binding bandstructure at high temperature. Since we see experimental features primarily corresponding to the surface state scattering at low temperature, we concentrate on comparing the scattering expected from the surface states when they are non-overlapping with the bulk bands at the same energy. Shown in figure 5c,d are the expected QPI from the surface states in both phases at -60 meV, shown on the same intensity color scale. We note that in this calculation we have included a thermal broadening of 300K to the bandstructure of both phases, to fairly compare the expected QPI from both phases to room temperature experiments. We see that the Weyl semimetal phase shows a strong signature of the "wings" that we have previously described, and that these "wings" are clearly present even after thermal broadening to 300 K. On the other hand, the monoclinic phase bandstructure shows very weak structure at the wavevectors of the wings. This happens in our calculations because the length of the surface state that is non-overlapping with the bulk states is much shorter in the trivial phase when compared to the semimetal phases, and consequently its intensity in the QPI pattern is much weaker.

In order to visualize the difference between the low and high temperature surface state structure, we proceed to perform STM/STS imaging at room temperature. Such experiments are in general technically challenging, and to avoid problems of sample contamination the experiments are performed on a freshly cleaved crystal with the microscope maintained at room temperature. We are successfully able to obtain atomic resolution imaging and spectroscopy (see supplementary information, section V, for real space images and detailed spectroscopy results). To verify the capability of our instrument at room temperature, we also perform successful imaging experiments on the well-known Shockley surface state of gold (111) (see supplementary information, section VI for details of these measurements). In the case of 1T'-MoTe$_2$, we show in Figure 5f a Fourier transform in the same energy range where QPI is observed at low temperature. It is evident from the data that the high temperature QPI shows no evidence for surface state dispersion at the wavevectors where it is observed at low temperature. We have already discussed the role of thermal broadening from the theoretical perspective above. We can also ask the purely experimental question – given the observed low-temperature QPI patterns in experiment shown in figures 1,3 and 4, what do we expect room temperature thermal broadening to do to these QPI patterns? We address this in two different ways. First, we perform a simple test to simulate thermal broadening at low temperature. The QPI signal is the Fourier transform of the experimentally measured differential conductance. The differential conductance is measured with a lock-in amplifier by applying an AC voltage and measuring the AC current response, and the overall broadening of the experimental spectrum is given by adding in quadrature the temperature broadening and the broadening due to the AC voltage. While we typically use an AC voltage that is comparable to the real sample temperature, we intentionally simulated the effect of high temperature by performing a low temperature measurement with a large AC voltage (RMS value of 30 meV), similar to room temperature. QPI results of this experiment at -80 meV are shown in Fig 5e (see supplementary information, section VII, for additional energies and more details on simulating thermal effects). As can be seen, we can still clearly see the "wing" features in the QPI along with additional features that are introduced after the broadening of the spectrum. A second test is to directly apply an energy broadening corresponding to room temperature to the measured low temperature QPI patterns. These calculations, described in supplementary information, section VII, also confirm that we expect the "wing" features to survive room temperature broadening. Both of these tests clearly show that the absence of QPI at high temperature is not an experimental artifact, but is rather a consequence of the difference in electronic properties of the sample at high temperature.

**Discussion:**

Our observation of the absence of the high temperature QPI at the wavevectors corresponding to the "wings" is reasonably consistent with our theoretical calculations, which show a heavy suppression of this feature in the trivial phase (figure 5d and supplementary information, section V). The complete absence of the feature at high temperature rather than a suppression of the intensity could additionally arise from a loss of topological protection of the surface state. Thus, at low temperature, the large surface state is topologically protected against the effects of disorder – at least at small momentum scattering, while no such topological protection – beyond weak antilocalization - should exist at high temperature. This would be compatible with a scenario where the surface states at high temperature are not observed due to disorder (as has been observed in other metallic surface states [71]) , while they survive at low temperature due to the topological protection. Future experiments on samples with varying levels of disorder can help shed light on whether surface state localization is significant at high temperature. We also note that this scenario where topological protection is responsible for explaining the low-temperature QPI is more compatible with the 4 Weyl point calculation rather than the 8 Weyl point calculation (where the large surface state is trivial at both low and high temperature). Regardless of the details of the high temperature surface state, this type of temperature-driven phase transition where the electronic structure is largely preserved but the topological character is changed presents a unique opportunity to isolate the effect of topological Fermi arcs on the properties of Weyl semimetals. By alloying (or by applying physical knobs such as strain or pressure) one could imagine pushing the phase boundary down to low temperature, where topological phenomena can be turned on and off by small changes in temperature. This offers the potential of isolating topological phenomena from non-topological, a major challenge in current spectroscopic and transport experiments.

**Methods:**

Samples were grown by flux-method. MoTe$_2$ powder was well mixed with sodium chloride (NaCl, molar ratio is about 1:7) and put in an alumina tube. This alumina tube was then sealed in a vacuumed quartz tube of pressure 0.18 Pa. Then, the glass tube was put in a Muffle furnace and heated at 1100 °C for 12 hours, followed by cooling to 900 °C at a rate of 0.5 °C/h. The quartz tube was then water quenched to room temperature to achieve the 1T' phase MoTe$_2$ crystals. Electrical resistivity measurement was carried out using a standard 4-probe technique in a Physical Property Measurement System (Quantum Design, 9T-PPMS).

Our DFT calculations are generally based on the Vienna ab initio simulation package [72], and use the core-electron projector augmented wave basis sets [73] with the generalized-gradient method [74]. Spin-orbital coupling is included self-consistently. The cutoff energy for wave-function expansion is 300 eV. Experimental lattice parameters are used throughout our calculations. To reduce the computation load in certain cases (e.g. in the comparison of the low-temperature and the high-temperature structures of MoTe$_2$), we construct tight-binding models by using the maximally localized Wannier function approach [75] and by keeping only the degrees of freedom corresponding to the Mo 4d orbitals and the Te 5p orbitals.

To obtain the spectral and spin densities on the surface of MoTe$_2$ from DFT calculations, we use a slab model of 4 surface unit cells (with 2 atomic layers per unit cell) in thickness, and with a (001) surface orientation. We use in-plane k-point grids of size 14x8 for the charge self-consistent calculations, and of size 1000x400 for the Fermi surface calculations. To obtain the surface spectral and spin densities from tight-binding models, we use the algorithm by Lopez Sancho et al. [76] to calculate the surface Green functions with 400x400 in-plane k-point grids.

The calculated spectral density is defined by $\rho_0(k, E) = Tr(A(k, E))$, and the spin densities are defined by $\rho_i(k, E) = Tr(\sigma_i A(k, E))$, with $A(k, E)$ the matrix spectral function and $\sigma_i$ the Pauli matrices for spin. The matrix spectral function $A(k, E)$ can be constructed from the Bloch eigenstates $\{\psi_n(k)\}$ (n is the band index), obtained from the DFT calculations, for a specific energy E, by taking its standard definition

$$A(k, E) = \sum_n -\frac{1}{\pi} Im\left(\frac{1}{E - E_n(k) + i\eta}\right) \psi_n(k) \psi_n^\dagger(k) \tag{2}$$

Here $E_n(k)$ is the energy of the n-th Bloch band, the eigenstate $\psi_n(k)$ is a column vector, and $\eta$ is a small number typically of value 2 meV. Alternatively, $A(k, E)$ can be obtained from the retarded surface Green functions $G^R(k, E)$, as in the case of tight-binding calculations, with an equivalent definition:

$$A(k, E) = \frac{i}{2\pi}\left[G^R(k, E) - G^R(k, E)^\dagger\right] \tag{3}$$

All $\rho_0$ and $\rho_{1,2,3}$ can be projected to specific layers by keeping $A(k, E)$ only for the appropriate atoms.

All angle resolved photoemission measurements were performed at the ANTARES [78] beamline located at the SOLEIL synchrotron, Gif sur Yvette, France. The beam spot size was 120 μm. The angular and energy resolution of the beamline at a photon energy of 18.8 eV are 0.2° and 10 meV, respectively. Linear polarized light was used. All data shown here were obtained at 90 K.

**Data Availability Statement:**

All relevant data is available upon request from the corresponding author.

**Acknowledgements:**

This work is supported by the National Science Foundation (NSF) via the Materials Research Science and Engineering Center at Columbia University (grant DMR 1420634), by grant DMR- 1610110 (A.N.) and by the Office of Naval Research grant number N00014-14-1-0501 (E.A.). Equipment support is provided by the Air Force Office of Scientific Research (grant number FA9550-16-1-0601) and FA9550-16-1-0031, J.P.). BAB acknowledges support from NSF EAGER Award NOA - AWD1004957, ONR - N00014-14-1-0330, ARO MURI W911NF-12-1-0461, NSF-MRSEC DMR-1420541. ZJW's work was supported by Department of Energy de-sc0016239, Simons Investigator Award, Packard Foundation and Schmidt Fund for Innovative Research. The work at Postech was supported by the Max Planck POSTECH/KOREA Research Initiative Program through National Foundation of Korea (NRF) funded by the Ministry of Science, ICT and Future Planning (No. 2016K1A4A4A01922028). The work at Rutgers was funded by the Gordon and Betty Moore Foundation's EPiOS Initiative through Grant GBMF4413 to the Rutgers Center for Emergent Materials. N.Z. acknowledges support by the NSF MRSEC program through Columbia in the Center for Precision Assembly of Superstratic and Superatomic Solids (DMR-1420634). MCA, JA and CC acknowledge Synchrotron SOLEIL, which is supported by the Centre National de la Recherche Scientifique (CNRS) and the Commissariat a` l'Energie Atomique et aux Energies Alternatives (CEA), France.

**Competing Interests:** None

**Author Contributions:**

ANB and EA performed STM experiments and analysis. JL, ZW and BAB performed theoretical analysis. NZ, JA, CC and MCA performed ARPES experiments and analysis. LZ and JK grew single crystals. BAB, S-WC and ANP advised. ANP is the guarantor of the research. ANB and APN co-wrote the paper, and all authors contributed to the discussion and preparation of the manuscript.

**Figures:**

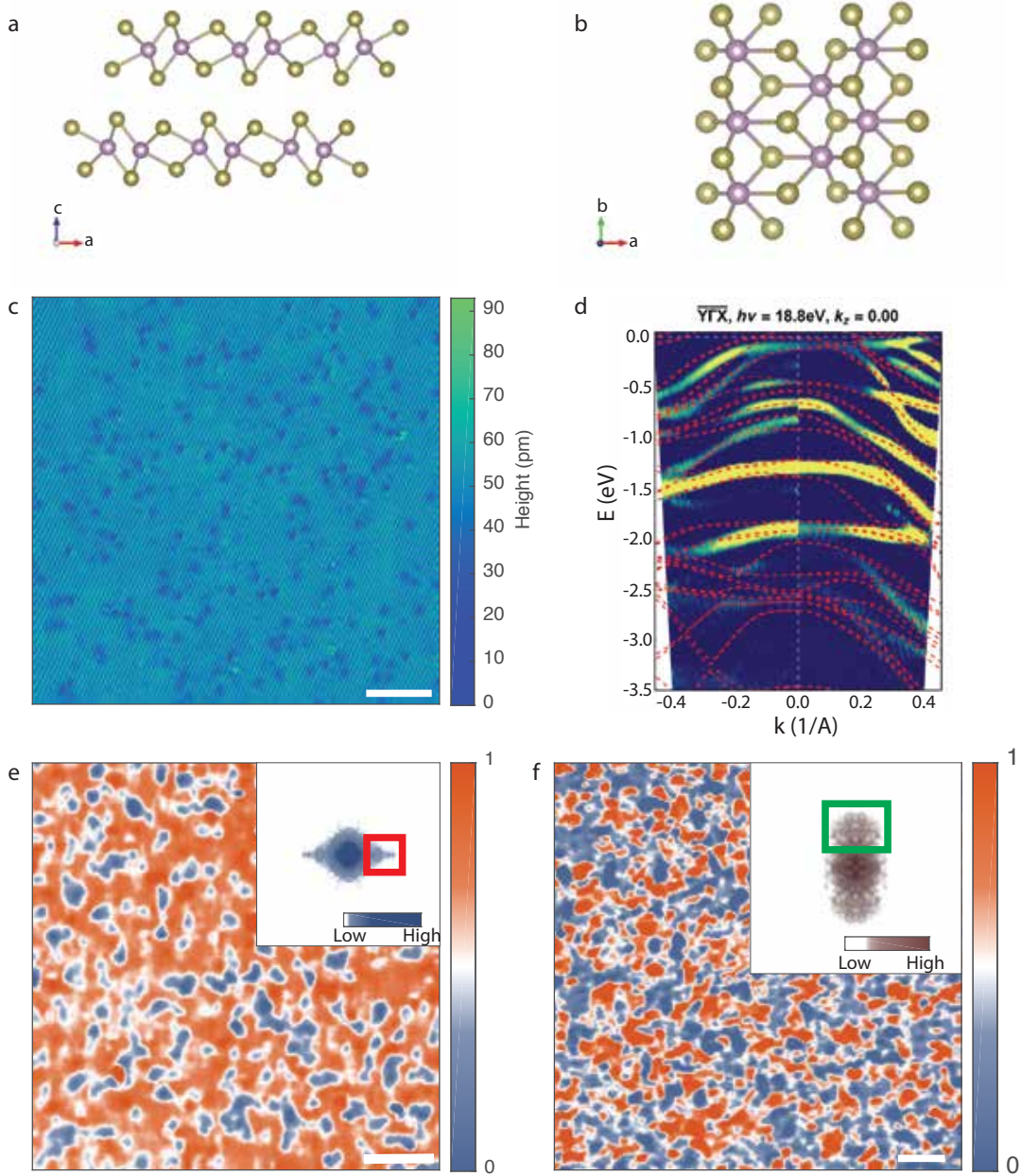

Figure 1 - Td-MoTe$_2$ (a,b) Ball and Stick models showing projected views of Td-MoTe$_2$. The Te atoms are represented in yellow, and the Mo atoms are in purple. (c) Constant-current STM topographic image of MoTe$_2$ (V = -200 mV, I = -150 pA, T = 6K). The rows seen run parallel to the b-axis of the crystal and come from the c-axis buckling in the Td crystal structure. (d) Two-dimensional cuts of the band structure along the high symmetry directions X-Gamma-Y over a wide range of energy as measured by ARPES (2[nd]

derivative applied) at T~90K. The red dashed lines represent theoretical predictions of the bulk band structure. (e,f) Differential conductance (dI/dV) map with insets showing Fourier transforms of the real space maps. (e) shows a map at -35 meV (normalization V = -300mV, I = -350pA, T = 6K). The Fourier transform inset shows wings aligned with the a axis of the crystal (red box). (f) shows a map of a different area at 50 meV (normalization V = 200 mV, I = 150 pA, T = 6K). The Fourier transform inset shows a vertically dispersing feature aligned with the b crystal axis (green box). All scale bars shown in this figure represent 10 nm.

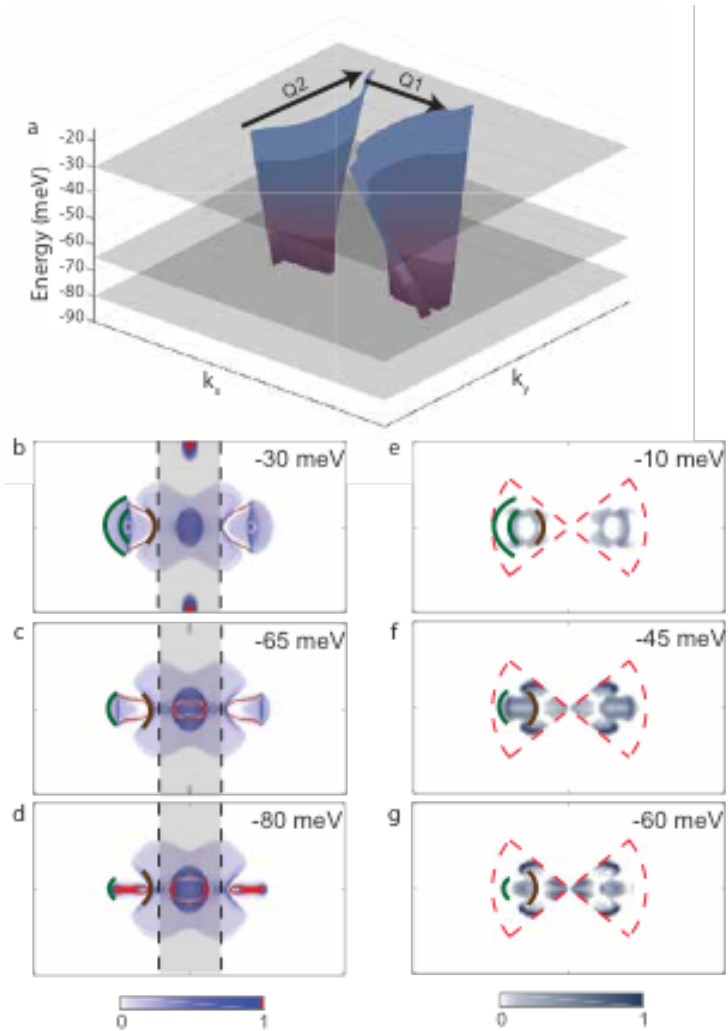

Figure 2 – Surface bands in Td-MoTe$_2$ (a) Dispersion associated with the surface bands. When the surface bands emerge at -80 meV, their Fermi surfaces are long and narrow, and as the energy increases, the Fermi surfaces open up and become shorter. The surface band Fermi surfaces persist as easily distinguishable features until approximately -10 meV. (b,c,d) Calculated constant energy contours, and (e,f,g) corresponding intensity maps (2$^{nd}$ derivative) measured through ARPES at various energies. The color map chosen for the calculated Fermi surface uses red for the most intense values, and blue for the rest of the band structure. This was deliberately chosen to highlight the fact that the surface states are much more intense than the bulk bands. The intensity corresponds to the density of states at a given momentum value, so the surface states by their two dimensional nature are very sharp and concentrated in momentum space in comparison with the projected bulk states. The shaded grey rectangle shows where a mask was applied to exclude the more intense bulk bands in the center of the BZ from the thresholding. The red dashed lines in (e,f,g) denote the region over which ARPES measurements were performed. The ARPES measurements of the Fermi surface (figure 2 e,f,g) show a surface bandstructure consistent with previous measurements. We outline the edges of the hole pocket in green and the edge of the electron pocket in brown to show how the evolution of the pockets matches between our ARPES data and calculated bandstructure. The data shown in figure e,f,g has been mirrored in $k_x$ and $k_y$ to show all quarters of the BZ. All of the bandstructure plots shown in this figure are cropped to the range $k_x$ = (-0.5π/a, 0.5π/a) and $k_y$ = (-0.5π/b, 0.5π/b).

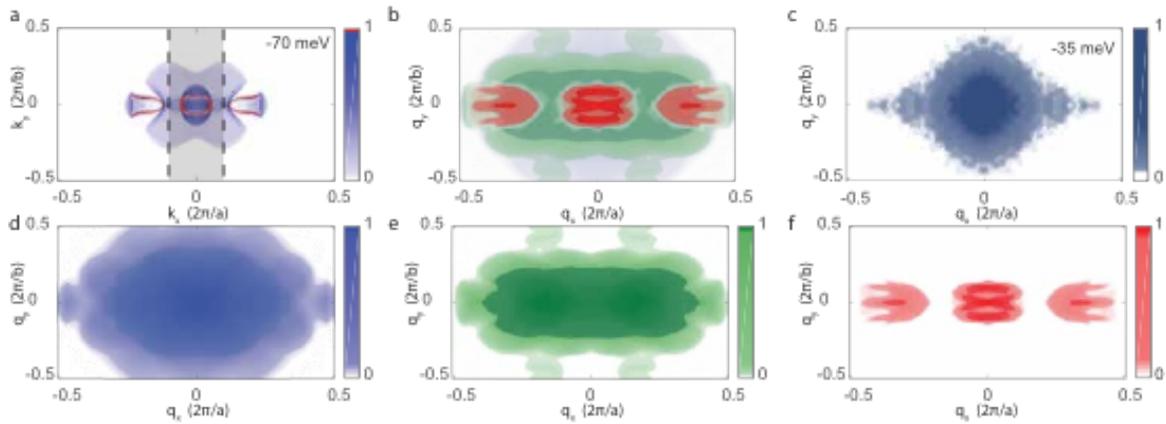

Figure 3 - Demonstration of the procedure used to separate QPI from bulk, surface and joint states. (a) shows the Fermi surface at -70 meV. Because the Fermi arcs sharply contrast with the bulk states, they are easily separated from the rest of the band structure through intensity thresholding. The grey rectangle shows where a mask is applied to exclude the more intense bulk bands in the center from the threshold. (b) shows all three QPI components overlaid, with bulk shown in blue, joint shown in green, and surface shown in red. (c) shows the Fourier transform of a differential conductance map at -35 meV, cropped to the first BZ for comparison to the calculated QPI. The horizontal wing feature matches with the QPI calculated from the Fermi arc surface states. (d) shows in blue the QPI coming from the bulk states only, with the Fermi arcs removed. (e) shows the joint QPI which comes from scattering between bulk and surface states. (f) shows the QPI coming only from the Fermi arc surface states.

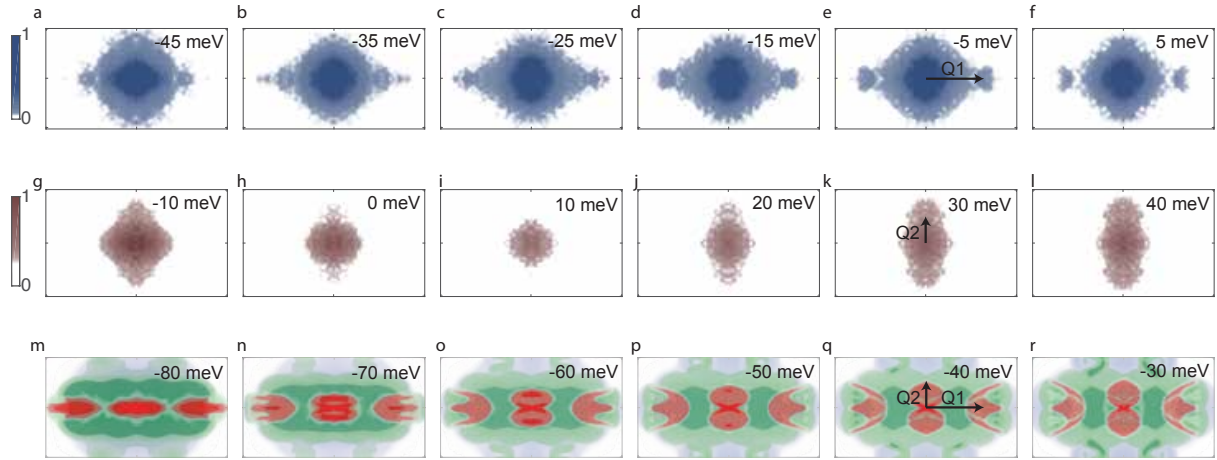

Figure 4 - Comparison of QPI from differential conductance maps and calculated QPI. (a-f) dI/dV map optimized for showing the horizontal dispersion. (g-l) dI/dV map optimized for showing the vertical dispersion. (m-r) QPI calculated from tight binding data using the procedure shown in figure 3. Blue represents QPI from bulk states, green is QPI from scattering between bulk and surface states, and red is QPI from surface states. All QPI images are cropped to the range $q_x$ = (-0.5π/a, 0.5π/a) and $q_y$ = (-0.5π/b, 0.5π/b), and aligned in energy as shown in supplementary.

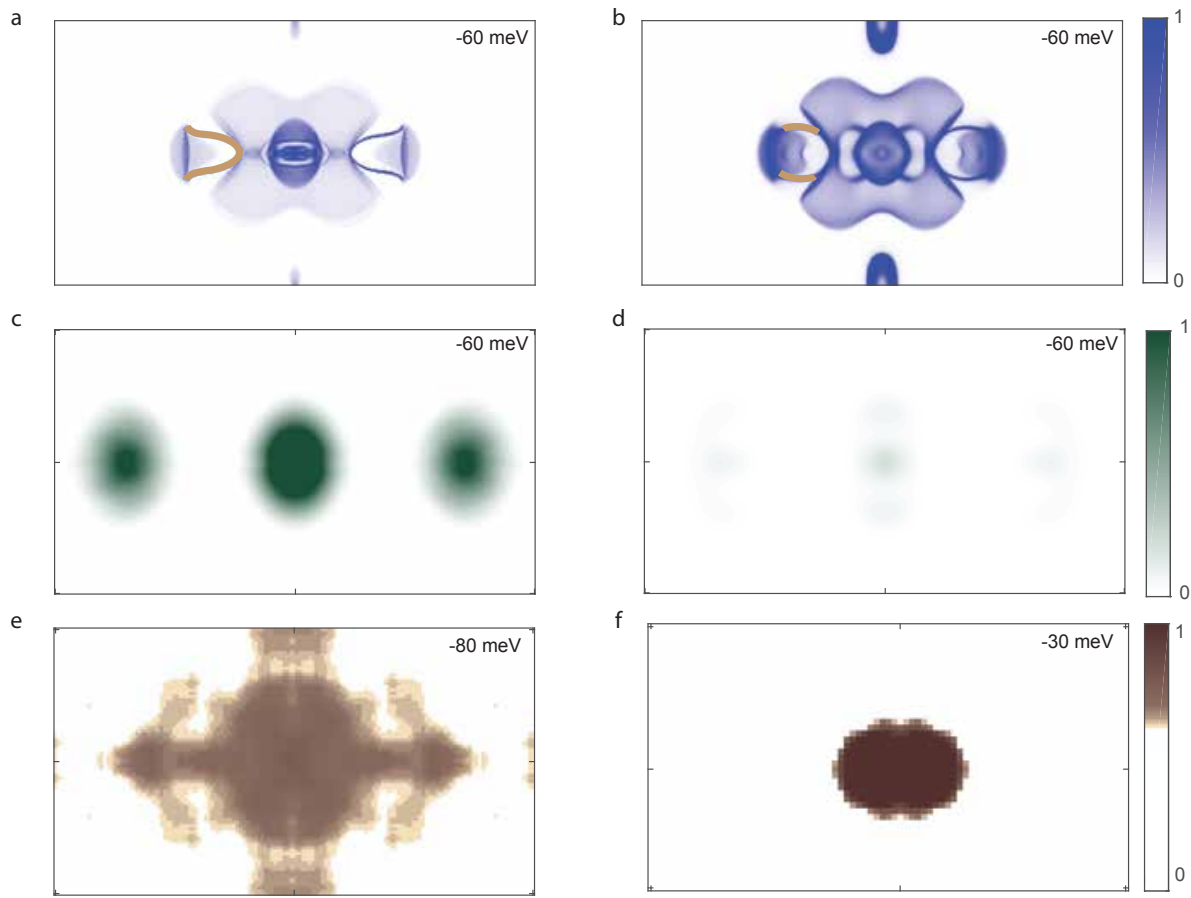

Figure 5 - Comparison of the high and low temperature phases. (a,b) The Fermi surfaces, calculated by tight binding, of the orthorhombic, Td phase at low temperature (a), and the monoclinic 1T' phase at high temperature. Surface states are present in both phases, highlighted on the left half of the brilliouin zone with brown contours. (c,d) QPI from tight binding surface states (only the parts that do not overlap with bulk bands) that have been thermally broadened to T=300K, as described in Supplementary Section VI, for the orthorhombic (c), and monoclinic (d) phases. Both phases show a horizontal wing feature in the QPI, but the intensity is much lower in the monoclinic phase, mostly due to overlap of the surface and bulk states. (g) Fourier transform of differential conductance map taken at low temperature with a 30 meV bias oscillation to simulate room temperature thermal broadening (see text). The horizontal wing feature is still visible in the QPI, even in the presence of energy broadening. (h) Fourier transform

of differential conductance map taken at room temperature. No features are seen in the QPI. . All of the plots shown are cropped to the first BZ, to the range $k_x$ or $q_x$ = (-0.5π/a, 0.5π/a) and $k_y$ or $q_y$ = (-0.5π/b, 0.5π/b).

# Supplementary Information For:

# Temperature-Driven Topological Transition in 1T'-MoTe$_2$

Ayelet Notis Berger[1,2]*, Erick Andrade[1]*, Alex Kerelsky[1], Drew Edelberg[1], Jian Li[3], Zhijun Wang[4], Lunyong Zhang[5,6], Jaewook Kim[6], Nader Zaki[7], Jose Avila[8], Chaoyu Chen[8], Maria C Asensio[8], Sang-Wook Cheong[6], Bogdan A. Bernevig[4,9,10,11], Abhay N. Pasupathy[1]

[1] Department of Physics, Columbia University, New York, NY 10027. [2] Department of Materials Science, Columbia University, New York, NY 10027. [3] Westlake Institute for Advanced Study, Hanzhou, China. [4] Department of Physics, Princeton University, Princeton, NJ 08520. [5] Laboratory for Pohang Emergent Materials & Max Plank POSTECH Center for Complex Phase Materials, Max Planck POSTECH/Korea Research Initiative , Pohang 790-784, Korea. [6] Rutgers Center for Emergent Materials & Department of Physics and Astronomy, Rutgers University, Piscataway, NJ 08854. [7] Department of Applied Physics & Applied Mathematics, Columbia University, New York, NY 10027. [8] Synchrotron SOLEIL Orme des Merisiers – Saint Aubin BP 48 – 91192 – GIF SUR YVETTE Cedex FRANCE [9] Laboratoire Pierre Aigrain, Ecole Normale Supérieure-PSL Research University, CNRS, Université Pierre et Marie Curie-Sorbonne Universités, Université Paris Diderot-Sorbonne Parice Cité, 24 rue Lhomond, 75231 Paris Cedex 05, France† [10] Donostia International Physics Center, P. Manuel de Lardizabal 4, 20018 Donostia-San Sebastián, Spain.† [11] Sorbonne Universités, UPMC Univ Paris 06, UMR 7589, LPTHE, F-75005, Paris, France. †

* These authors contributed equally to this work.

†On sabbatical

## I. Resistivity Measurements of Samples Used

Figure S1 shows the temperature dependence of resistivity in the 1T'-MoTe$_2$ bulk crystal. The sample was grown using the flux method (as described in Methods). Electrical resistivity measurement was carried out using a standard 4-probe technique in a Physical Property Measurement System (Quantum Design, 9T-PPMS).

## II. Data Processing

Figure S2 shows the typical steps followed for processing the QPI from differential conductance maps shown in this paper. We first take the fourier transform of the conductance map at each energy and visualize it with a linear colormap (S2 a-f) . The "wings" that come from the surface state scattering are clearly visible in this raw Fourier transform. To increase the signal to noise ratio in the image, we mirror symmetrize the data along the $k_x$ and $k_y$ axes. The symmetrized data with two different colormaps is shown in Figure S2 g-l and Figure S2 m-r (the one chosen in the main text). No additional QPI features are generated as a consequence of symmetrization procedures in the Fourier transform.

## III. Spectroscopy maps at low temperature

We present two spectroscopy maps in figure 4 of the main text at low temperature to better emphasize the two main features of the QPI (inter-arc and intra-arc scattering). To understand the reason we choose maps of different sizes to display the horizontal and vertical dispersions clearly, we need to consider several factors of the experiment. The first issue is the time required for data acquisition. At the current resolution (256 pix), data acquisition for a map takes 1-1.5 days, which is the limit at which we can control drift and operate without cryogen refills at low temperature. For a given pixel resolution, increasing the spatial size of the map results in a higher Fourier space resolution. However, this comes at a price, since a larger map averages over more disorder in the sample. The presence of this disorder creates an overall near-Gaussian background centered at the origin in Fourier space, the size and intensity of which is sample dependent. We optimize the size of the map to visualize a given QPI feature keeping these constraints in mind. We have seen the main QPI features presented in the manuscript in over a dozen maps taken on different areas of the samples, ensuring that it is not an artifact related to disorder realization.

Figure S3 shows the real space images (S3a-f) and Fourier transforms (S3g-l) for several energies of a differential conductance map. It also shows the Fourier transforms cropped to the first Brillouin zone for comparison to other data sets (S3m-r). The area of the map is chosen to emphasize larger momentum transfer, clearly showing the wings aligned with the a-axis of the crystal. This feature comes from inter-arc scattering. S4 shows another differential spectroscopy map, chosen to show a larger real space image (S4a-i), in order to focus on smaller momentum scattering. This map emphasizes the vertically dispersing feature which is related to intra-arc scattering. S4j-r show the Fourier transform of the real space spectroscopy map, and S4s-aa show the QPI cropped to the first Brillouin zone for comparison to other data sets. The QPI cropped to the first Brillouin zone has been symmetrized as discussed in section II above.

### IV. Energy alignment between spectroscopy maps taken at different times

Our maps are obtained on several locations of multiple samples that have been exposed to vacuum conditions for varying amounts of time (depending on the time to acquire data). Between the various experimental runs at different sample locations, we find Fermi level shifts from run to run that are in the range of tens of meV. Such shifts have been observed in the past on surfaces due to sample inhomogeneity and aging. Here we describe the procedure for aligning the features seen at different energies in the data sets shown in figure 4 of the main text. Figure S5a-f and S5g-l show the QPI cropped to the first Brillouin zone from the same spectroscopy map shown in figure S4. Different color maps are used to emphasize the vertical dispersion (S5a-f) and the horizontal wing feature (S5g-l). Using the color map that emphasizes the horizontal wing feature, we align this data set with that shown in figures 1e, S3, and S5m-r. We find that the data set from S4 is offset by +35 meV from the S3 data set. We also match the strong horizontal wing feature with the Fermi arc surface state QPI predicted from the tight binding model, shown in S5s-x. We find that the S3 data set is offset by 35 meV from the tight binding calculations and the S4 data set is offset by 70 meV from the tight binding calculations. Similar shifts are seen in over a dozen maps taken on different areas of the sample over a period of approximately six months.

### V. Room Temperature Experiments

Figure S6 shows atomic resolution topographies and spectroscopy maps taken at room temperature, in the monoclinic phase. S6a,b show atomic resolution STM topographies. The lattice constants match those measured for this phase [7]. S6c shows the fourier transform of a differential conductance map at 0 mV. Sd-i show the real space data from various energies of a differential conductance map, and the insets show the fourier transform at each energy. No strong features are shown which could be attributed to QPI from the trivial surface states in the non-topological phase. All data shown here has been drift corrected and Gaussian smoothed ($\sigma$=2 pixels).

### VI. Room Temperature QPI in Gold (111)

Figure S7 shows the fourier transform at several energies of a differential conductance map taken of the (111) surface of gold at room temperature. This shows the QPI from the well known surface state of gold (111) is present even in the presence of lattice vibrations and energy broadening at room temperature. The scattering vector $2k_f$ is indicated with a red arrow in S7f-j.

### VII. Simulated Thermal Broadening for Spectroscopy Maps in the Td Phase

We estimate the effect of thermal broadening on our low temperature spectra and maps in two separate ways as described below:

(a) <u>Fermi function broadening of low-temperature spectra:</u> In the weak tunneling regime of typical STM experiments, the current $I$ at position $r$ and voltage $V$ is given by

$$I(r,V,T) = C \int_{-\infty}^{\infty} \rho_s(r,E,T=0)[f(E+eV) - f(E)]dE$$

where C is a constant including the tip density of states and tunnel matrix element, $\rho_s(E)$ is the density of states of the sample, and $f(E)$ is the Fermi-Dirac distribution at a given temperature. This implies that the differential conductance is given by:

$$\frac{dI}{dV}(r,V,T) = C\int_{-\infty}^{\infty} \rho_S(r,E,T=0)\frac{df}{dV}(E+eV)dE = \int_{-\infty}^{\infty} \frac{dI}{dV}(r,E,T=0)\frac{df}{dV}(E+eV)dE$$

i.e., given the $T=0$ differential conductance $\frac{dI}{dV}(r,E,T=0)$, we can calculate the differential conductance at non-zero temperature $\frac{dI}{dV}(r,V,T)$ by simply convolving with the derivative of the Fermi-Dirac distribution at that temperature. We perform this procedure at every location in a map obtained at low temperature to generate maps that would correspond to a room temperature energy broadening. We then take the Fourier transform of these broadened maps to simulate the effect of room temperature energy broadening on the QPI patterns. The results of this procedure are shown in figure S10. Panels S10 a-e show the QPI patterns before room temperature broadening, while panels S10 f-j show the broadened QPI data. While some of the detail is missing from the thermally broadened data, the most distinctive feature of the QPI, the wings, are still clearly visible.

(b) <u>Energy broadening via a lock-in amplifier:</u> In a practical STM experiment, the differential conductance is obtained using a lock-in amplifer. The lock-in supplies a sinusoidal voltage $V_{ac} = \Delta V \sin(\omega t)$ which is added to the dc bias $V_{dc}$ at which the differential conductance is to be obtained. The total time dependent voltage applied to the junction is

$$V(t) = V_{dc} + \Delta V \sin(\omega t)$$

This results in a sinusoidal current response in the tunnel junction given by

$$I(t) = I_{dc} + \Delta I \sin(\omega t)$$

The ac amplitude of the current response $\Delta I$ is read by the lock-in amplifier. In linear response, the ratio $\Delta I/\Delta V$ is simply the differential conductance at voltage $V_{dc}$. From section VII, part a, above,

$$I(r,V,T=0) = C\int_0^{eV} \rho_S(r,E,T=0)dE$$

In linear response, this gives the usual relationship to the local density of states:

$$\frac{\Delta I}{\Delta V} \approx \frac{dI}{dV}(r,V_{dc},T=0) = C\rho_s(r,V_{dc},T=0)$$

In practice, an ac voltage $\Delta V$ that is smaller than the temperature is chosen to avoid energy broadening of the spectrum. However, we can choose to intentionally broaden our spectrum by applying a large ac voltage. In this case, we cannot use linear response, and instead we have to calculate the ac current response directly:

$$I(r,V(t),T=0) = I_{dc} + \Delta I \sin(\omega t) = C\int_0^{eV(t)} \rho_s(r,E,T=0)dE = C\int_0^{e[V_{dc}+\Delta V \sin(\omega t)]} \rho_s(r,E,T=0)dE$$

giving

$$\Delta I \sin(\omega t) = C\int_{eV_{dc}}^{e(V_{dc}+\Delta V \sin(\omega t))} \rho_s(r,E,T=0)dE$$

To extract the value of $\frac{\Delta I}{\Delta V}$, the lock in amplifier integrates the current over one time period of the sinusoidal modulation:

$$\int_0^\tau I(t)\sin(\omega t)dt = \int_0^\tau (I_{dc}+\Delta I \sin(\omega t))\sin(\omega t)dt = \pi\Delta I$$

Therefore:

$$\frac{\Delta I}{\Delta V} = \frac{C}{\pi\Delta V}\int_0^\tau \left(\int_0^{e(V_{dc}+\Delta V \sin(\omega t))} \rho_S(r,E,T=0)dE\right)\sin(\omega t)\,dt$$

We compare the effect of broadening the spectrum using a large ac voltage to the case of Fermi distribution broadening using a simple case, a dirac-delta density of states centered at E=0. Shown in Figure S8 is the comparison for a lock-in broadening of 30 mV RMS (figure S8a) and room temperature Fermi-Dirac function broadening (figure S8b). The full width at half maximum for the two cases is comparable (~90 mV for thermal broadening, and ~ 73 mV for lock-in-amplifier broadening). The horizontal QPI features persist over a large enough energy range that this broadening still preserves the features.

**VIII. Comparison of QPI Calculated from Tight Binding**

In figures S11, we show a detailed comparison of the QPI from the surface bands calculated by tight binding for the monoclinic and orthorhombic phases. As discussed in the main text, we select the surface states through thresholding. We additionally apply a mask to exclude bulk bands, and we calculate the QPI based on only scattering between surface states. Since we are comparing the QPI from these calculations to experimental data taken at room temperature, we apply thermal broadening of 300K to the surface bands before calculating the QPI. The method by which we apply thermal broadening is discussed above, section VII. While both phases show horizontal and vertical dispersing features in the QPI, the intensity of the QPI is much stronger in the orthorhombic phase, mostly due to the large overlap of the surface and bulk states in the monoclinic phase.

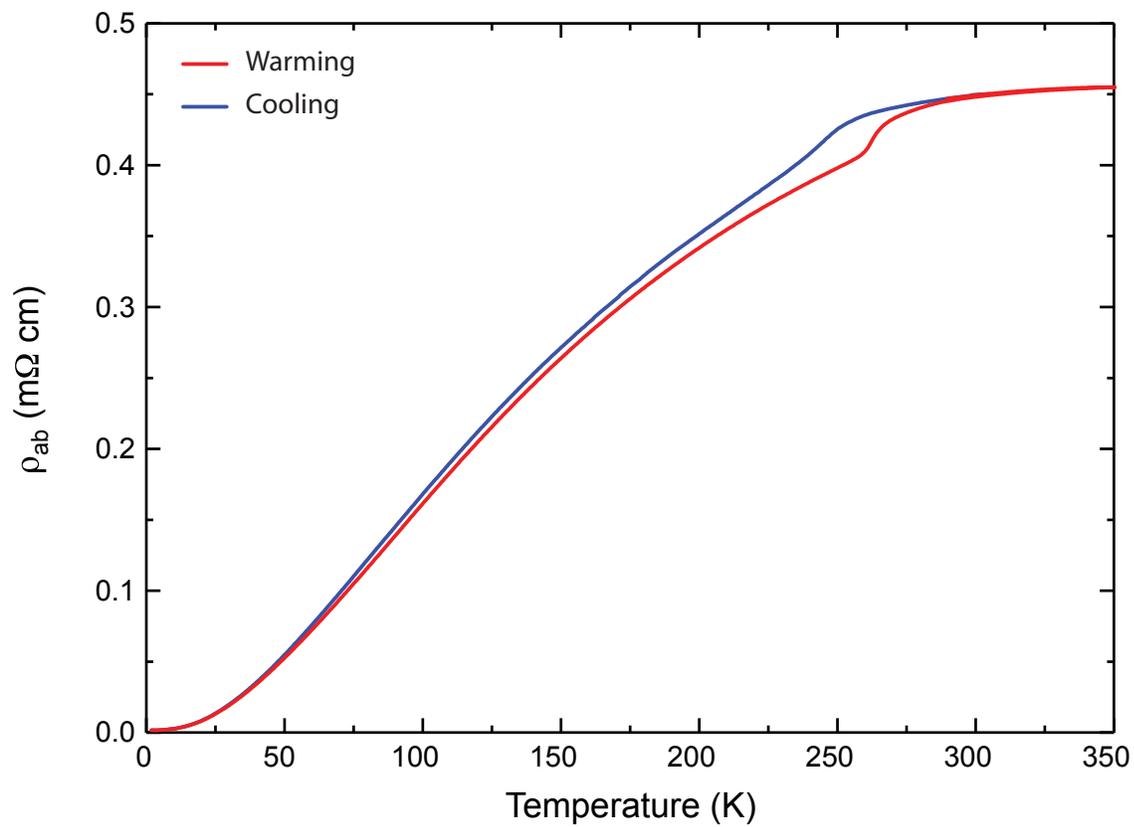

**Figure S1** – Resistivity measurement of 1T' MoTe$_2$ bulk crystal, taken using a standard 4-probe technique in a Physical Property Measurement System (Quantum Design, 9T-PPMS).

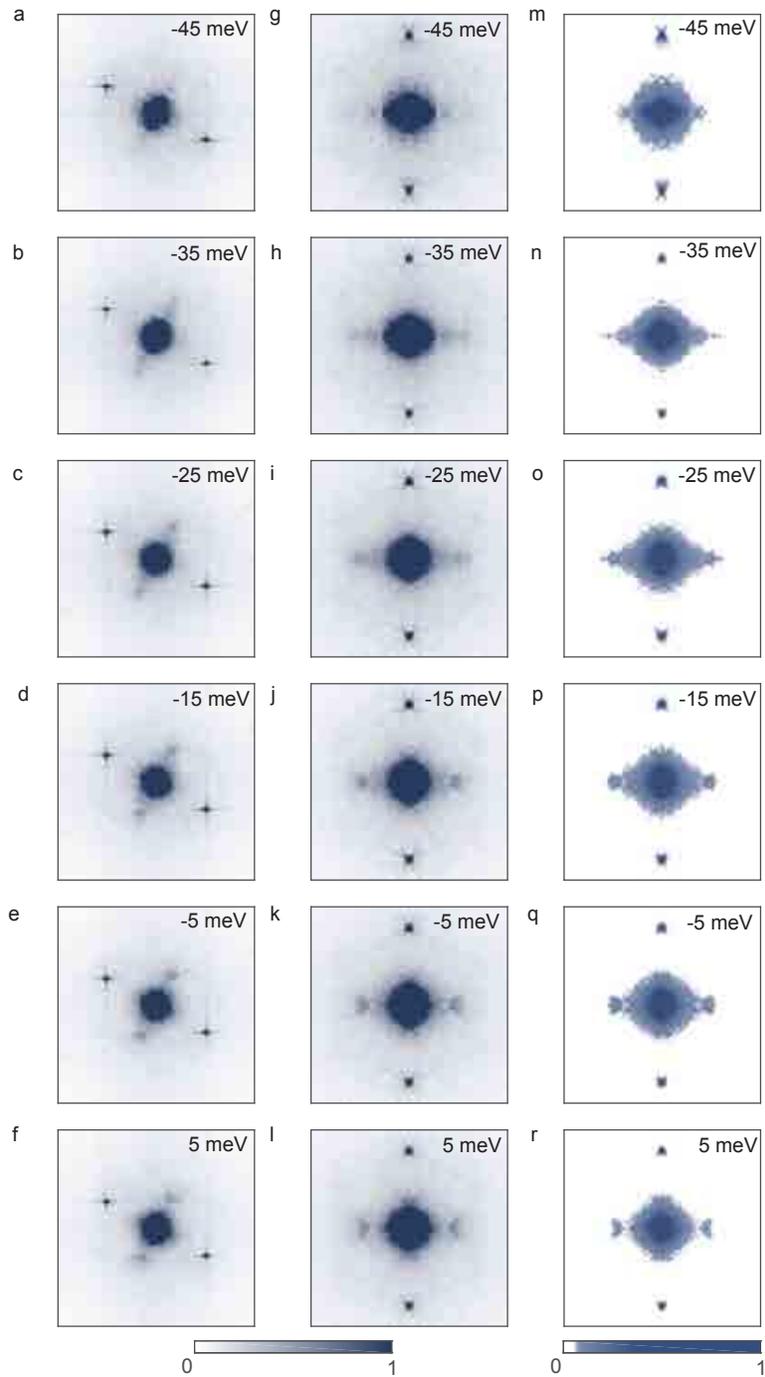

**Figure S2** – Step by step processing for QPI from a differential conductance map (V=-300 mV, I = -350 pA, T 6K). (a-f) are fourier transforms of the real space differential conductance at various energies. These fourier transforms are shown using a linear colormap. Wings from the QPI are visible but are slightly obscured by background noise. (g-l) show the same fourier transforms after they have been symmetrized. (m-r) show the same symmetrized data, but with a non-linear colormap that eliminates the background noise and makes the QPI more visible for illustration purposes.

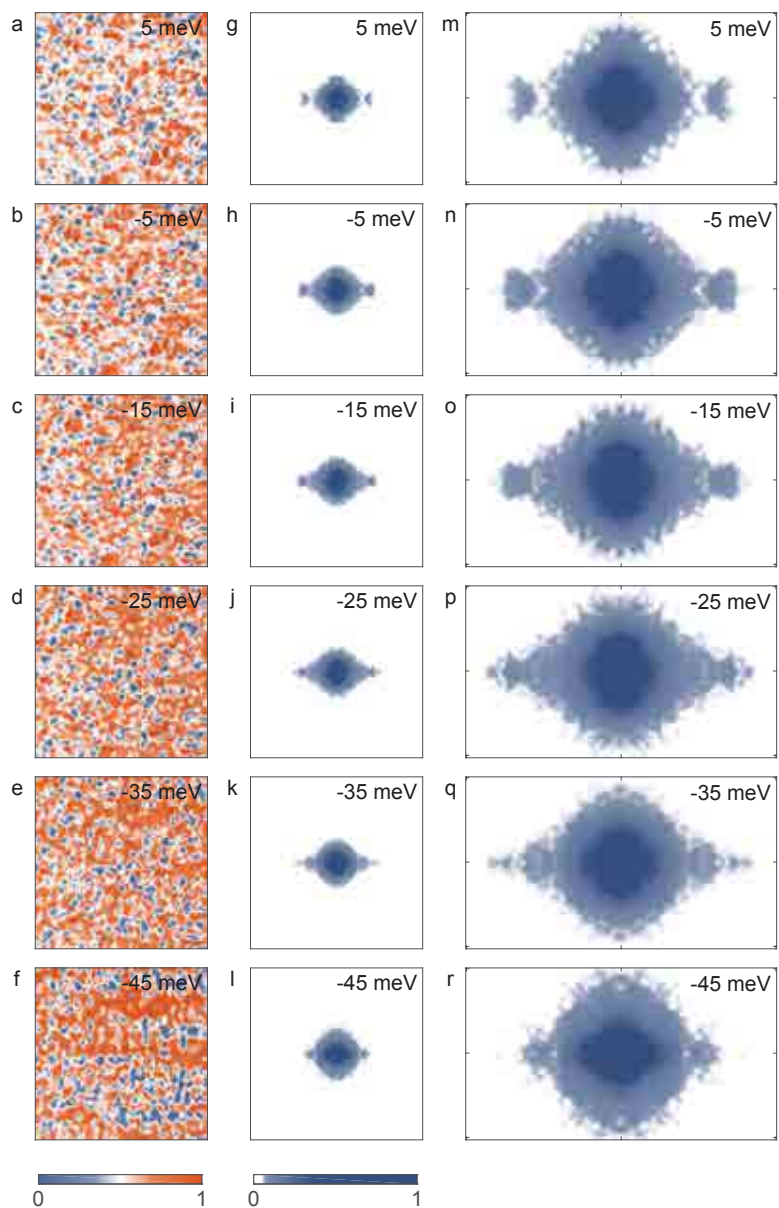

**Figure S3** - Differential conductance map showing the horizontal wing feature in the QPI. Normalization V = -300 mV, I = -350 pA, T= 6K. **(a-f)** show the real space data at various energies. The area shown is a square with side length 58.5 nm. **(g-l)** show the Fourier transform of the real space images shown in **(a-f)**. **(m-r)** show the Fourier transforms at each energy cropped to show only the first Brillouin zone.

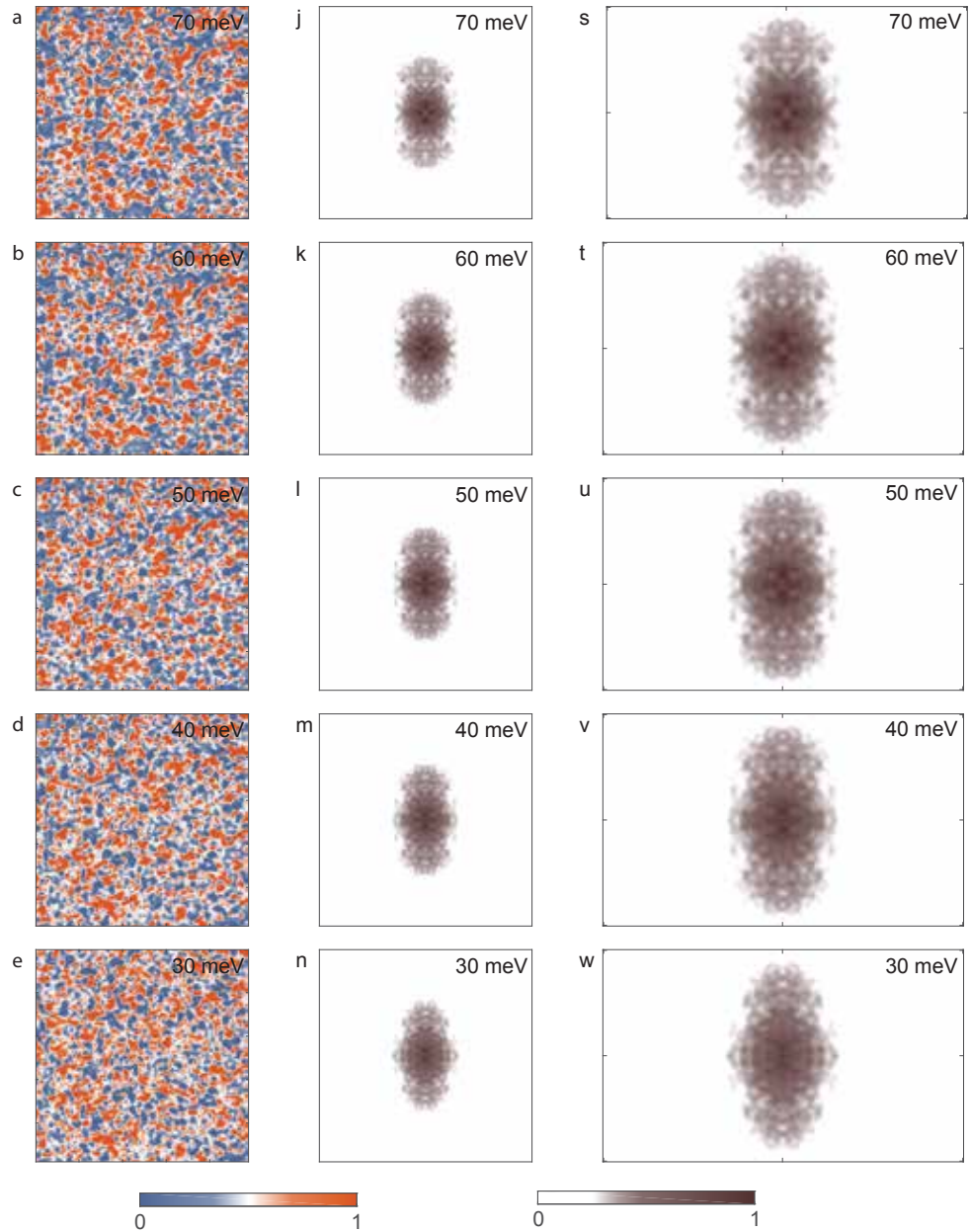

**Figure S4** (continued on next page) - Differential conductance map showing the vertically dispersion feature in the QPI. Normalization 200mV, I = 150 pA, T=6K. **(a-i)** show the real space data of the map at various energies. The area shown is square with side length 86.8 nm. **(j-r)** is the Fourier transform of the real space data at each energy, showing the QPI, and **(s-aa)** is the same Fourier transforms cropped to the first Brillouin zone.

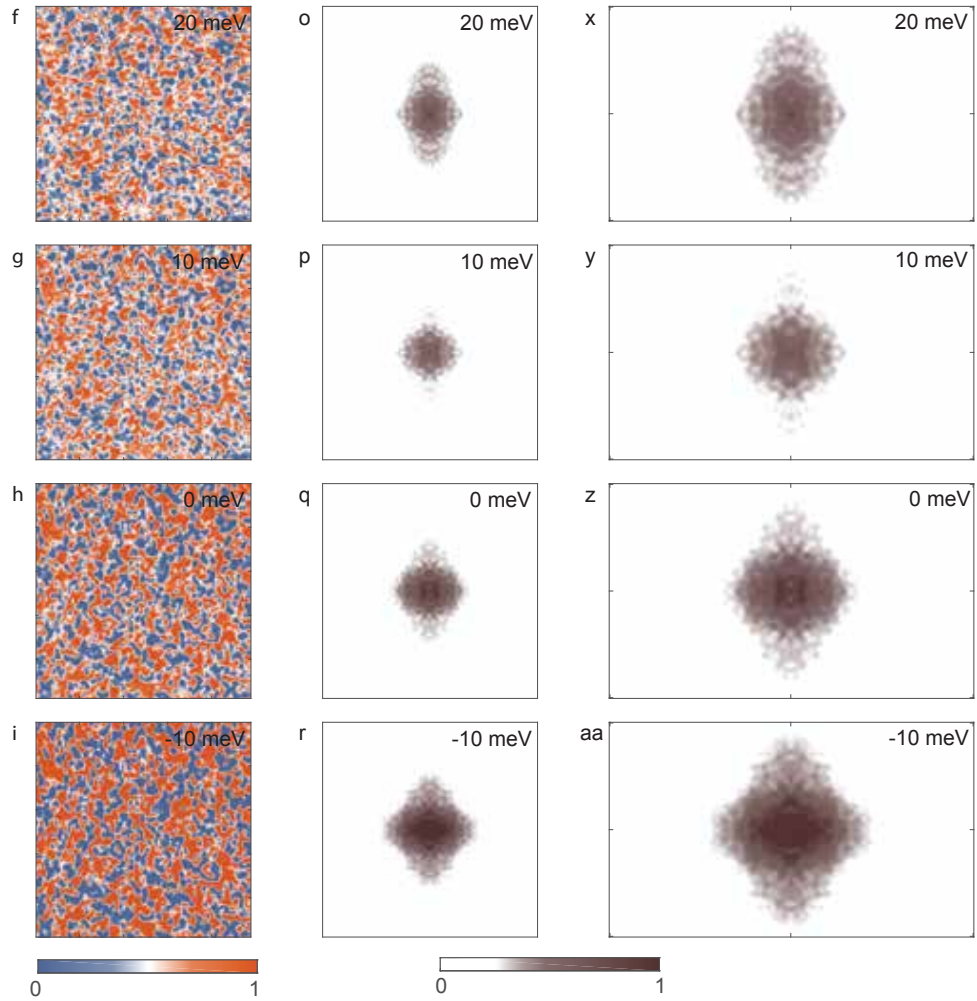

**Figure S4** (continued from previous page) - Differential conductance map showing the vertically dispersion feature in the QPI. Normalization 200mV, I = 150 pA, T=6K. **(a-i)** show the real space data of the map at various energies. The area shown is square with side length 86.8 nm. **(j-r)** is the Fourier transform of the real space data at each energy, showing the QPI, and **(s-aa)** is the same Fourier transforms cropped to the first Brillouin zone.

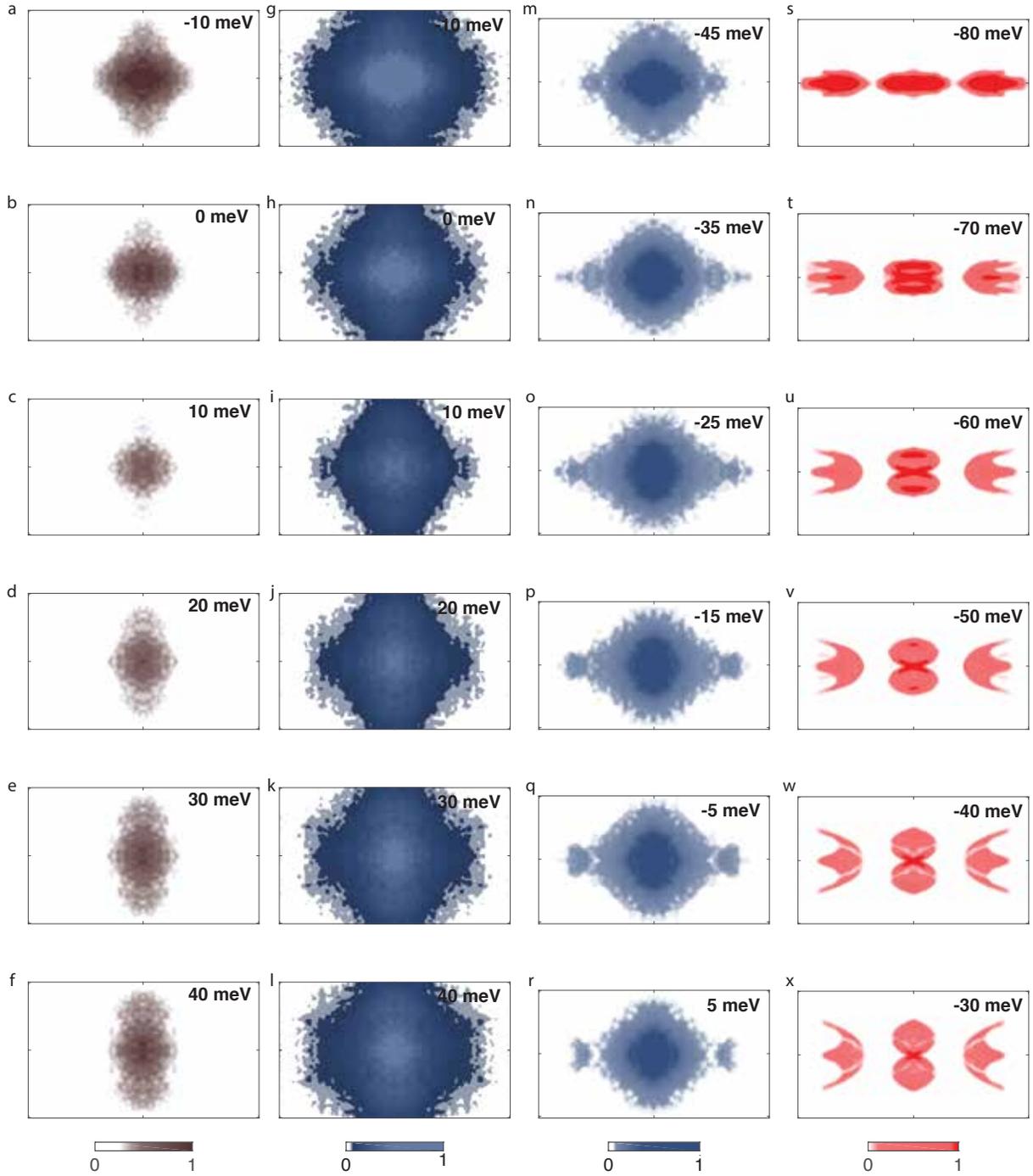

**Figure S5** - Energy alignment of the tight-binding calculations and the two low temperature data sets used in the paper. The Fourier transform of the differential conductance maps is cropped to the first Brillouin zone, to the range $q_x = (-0.5\pi/a, 0.5\pi/a)$ and $q_y = (-0.5\pi/b, 0.5\pi/b)$, for comparison to data of different scales. **(a-f)** and **(g-l)** show the same data set with different color maps. The color map in **(a-f)** emphasizes the vertically dispersing feature in the QPI, but the same data, viewed with a different color map shows the horizontal wing feature which is aligned in energy with the other main data set **(m-r)** and the tight binding QPI predictions **(s-x)**. This technique is used to determine the shifts in the Fermi level between data sets.

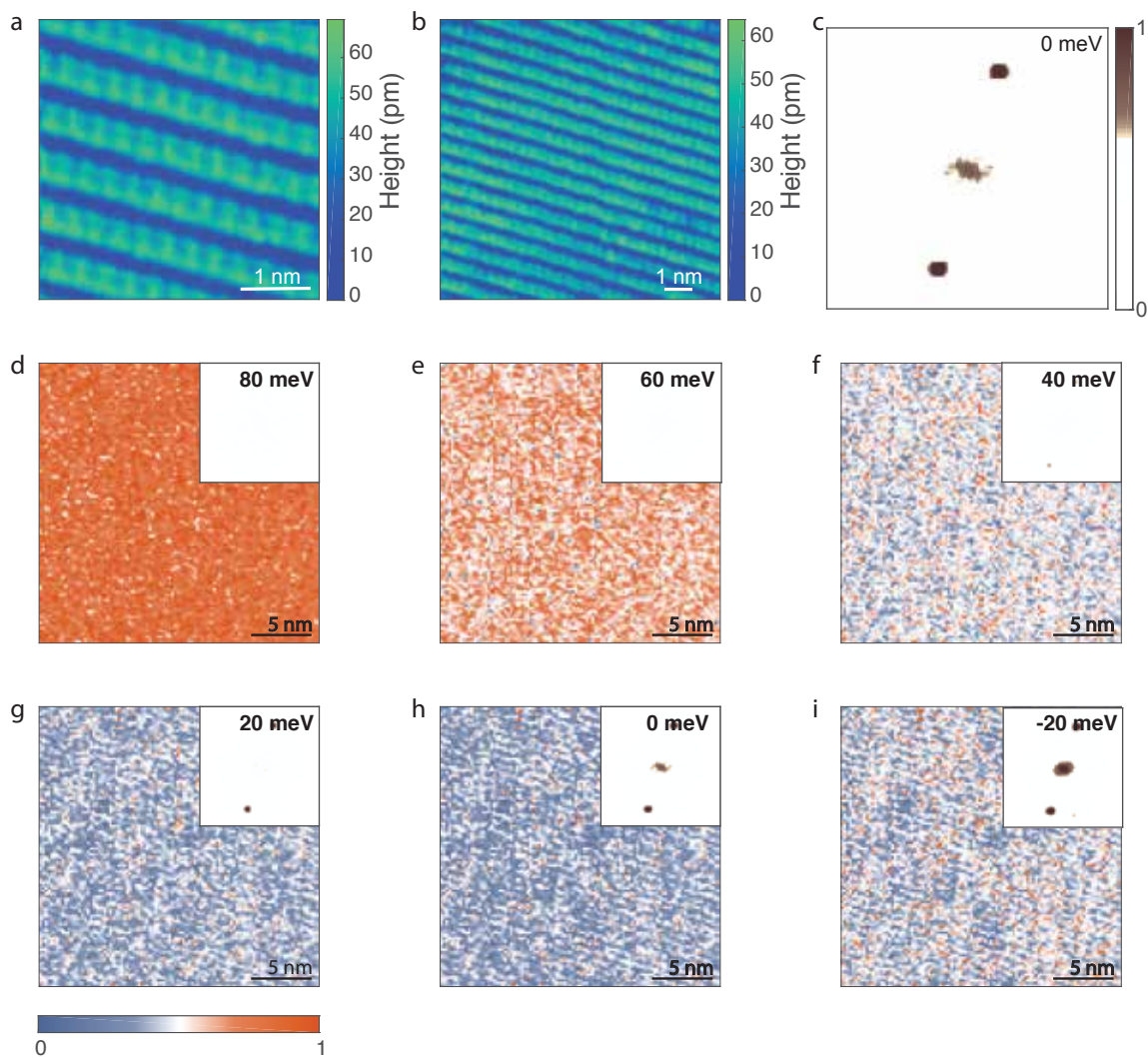

**Figure S6** - **(a,b)** Atomic resolution topographies of MoTe2 at room temperature (V = 20 mV, I = 3 nA). **(c)** Fourier transform of the room temperatures differential conductance map at 0 meV (real space shown in panel h). The peak in the center comes from disorder in the sample and is always present to some extent. The two peaks in the upper right and lower left are the periodic signal from atomic rows. No distinct features of QPI are seen in this data. The Fourier transforms of the map at other energies show no qualitative difference, so they are not shown here. **(d-i)** Real space differential conductance map of MoTe2 at room temperature (V = 200 mV, I = 100 pA), shown at different energies. The insets show the fourier transform at each energy.

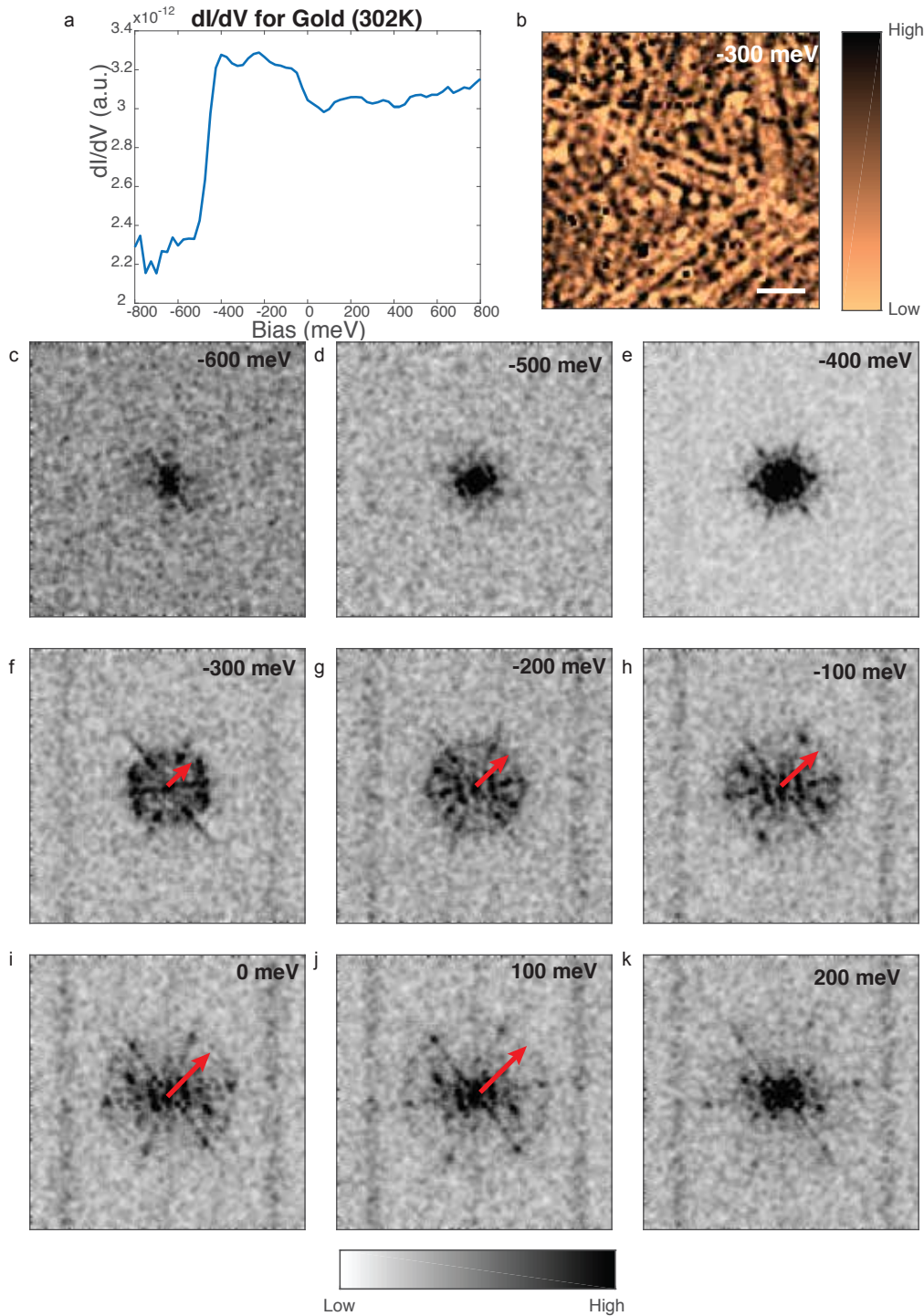

**Figure S7** – Differential conductance map of gold (111) at 302K (V = -800 mV, I = -100 pA). (a) shows the average spectrum over the entire surface covered in real space, showing the clear presence of a surface state around -450 mV. (b) shows the real space differential conductance at -300 mV. The scale bar is 10 nm. (c-k) show the fourier transform of the conductance at various energies (no further processing is applied). The scattering vector $2k_f$ is indicated with red arrows in f-j. The circle that emerges and grows with increasing energy reflects the expected dispersion of the well known gold (111) surface state. The surface state is not destroyed by lattice vibrations at room temperature.

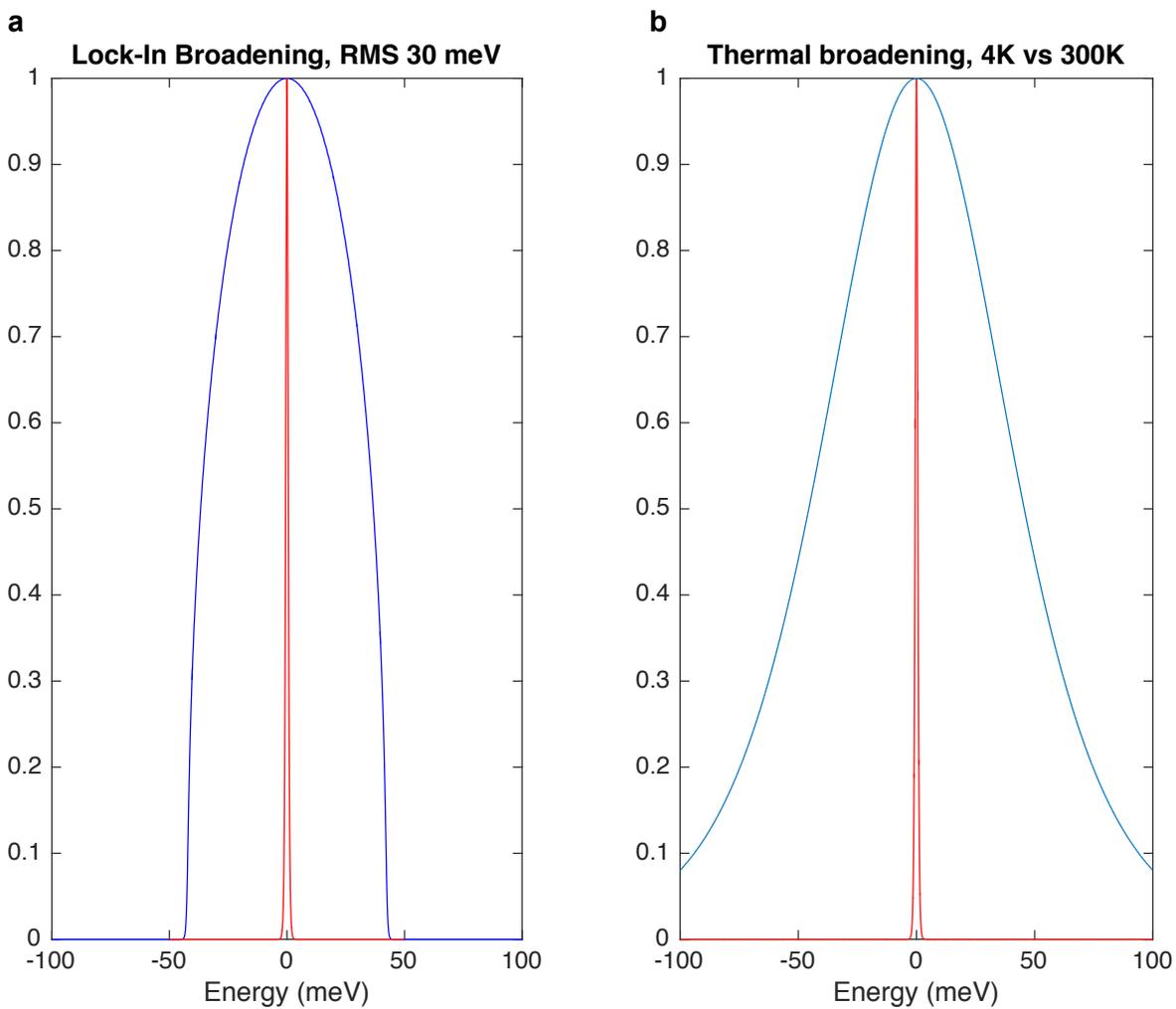

**Figure S8** - Comparison of energy broadening from a large bias modulation in the lock-in amplifier (a) and thermal broadening at 300K (b). The curve in red shows the derivative of the Fermi-Dirac distribution at 4K, and the curve in blue shows the differential conductance measured for a 30 mV bias modulation (a) and a temperature of 300K (b).

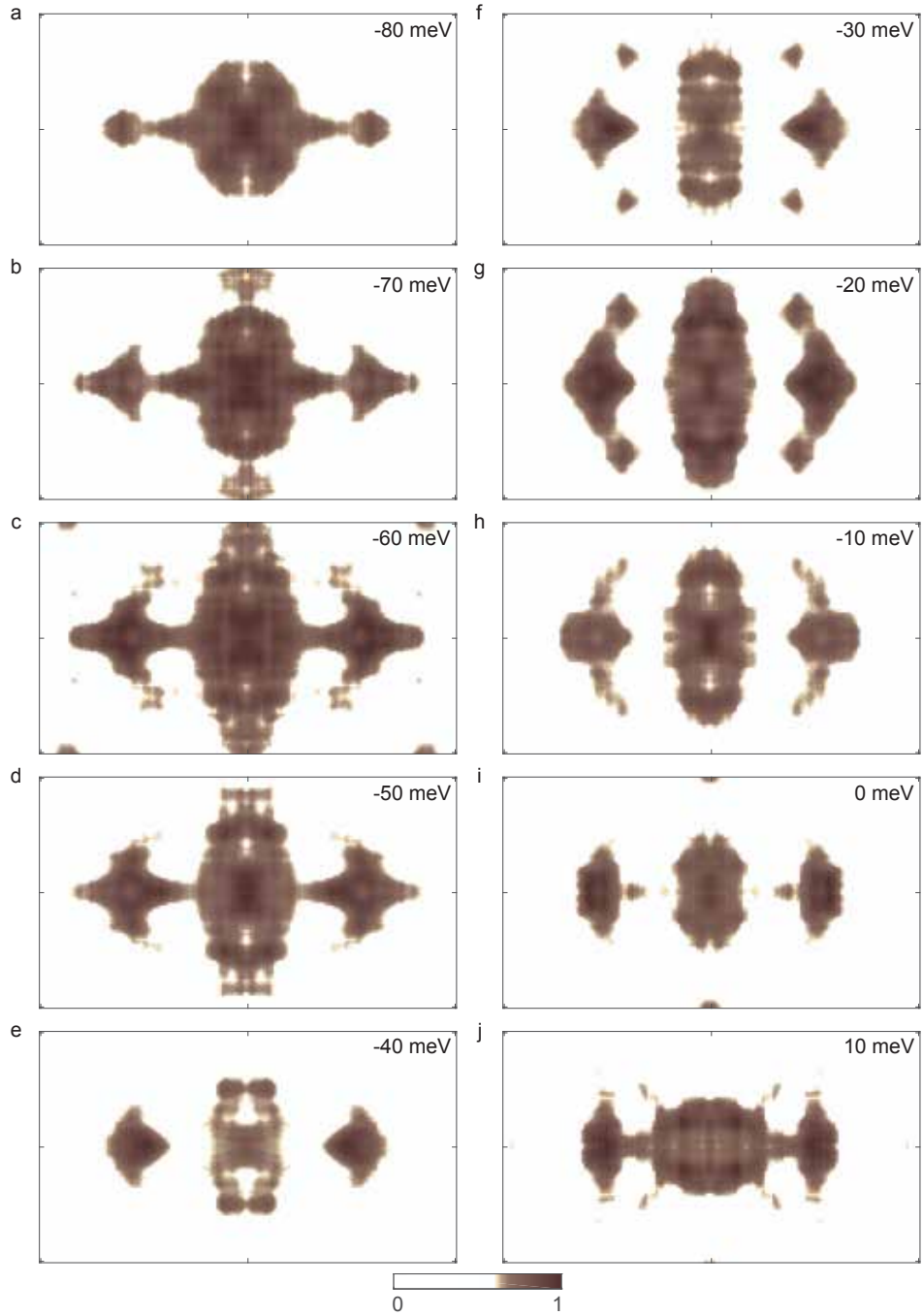

**Figure S9** - Fourier transform of differential conductance map (V = 180 mV, I = 220 pA). Data has been symmetrized and cropped to the first Brillouin zone, to the range $q_x$ = (-0.5π/a, 0.5π/a) and $q_y$ = (-0.5π/b, 0.5π/b). The data are taken at low temperature, in the Td phase, with a large bias oscillation to simulate thermal broadening that is expected in room temperature measurements. The strong wing features and vertical dispersion are still seen in the low temperature QPI despite the imposed energy broadening.

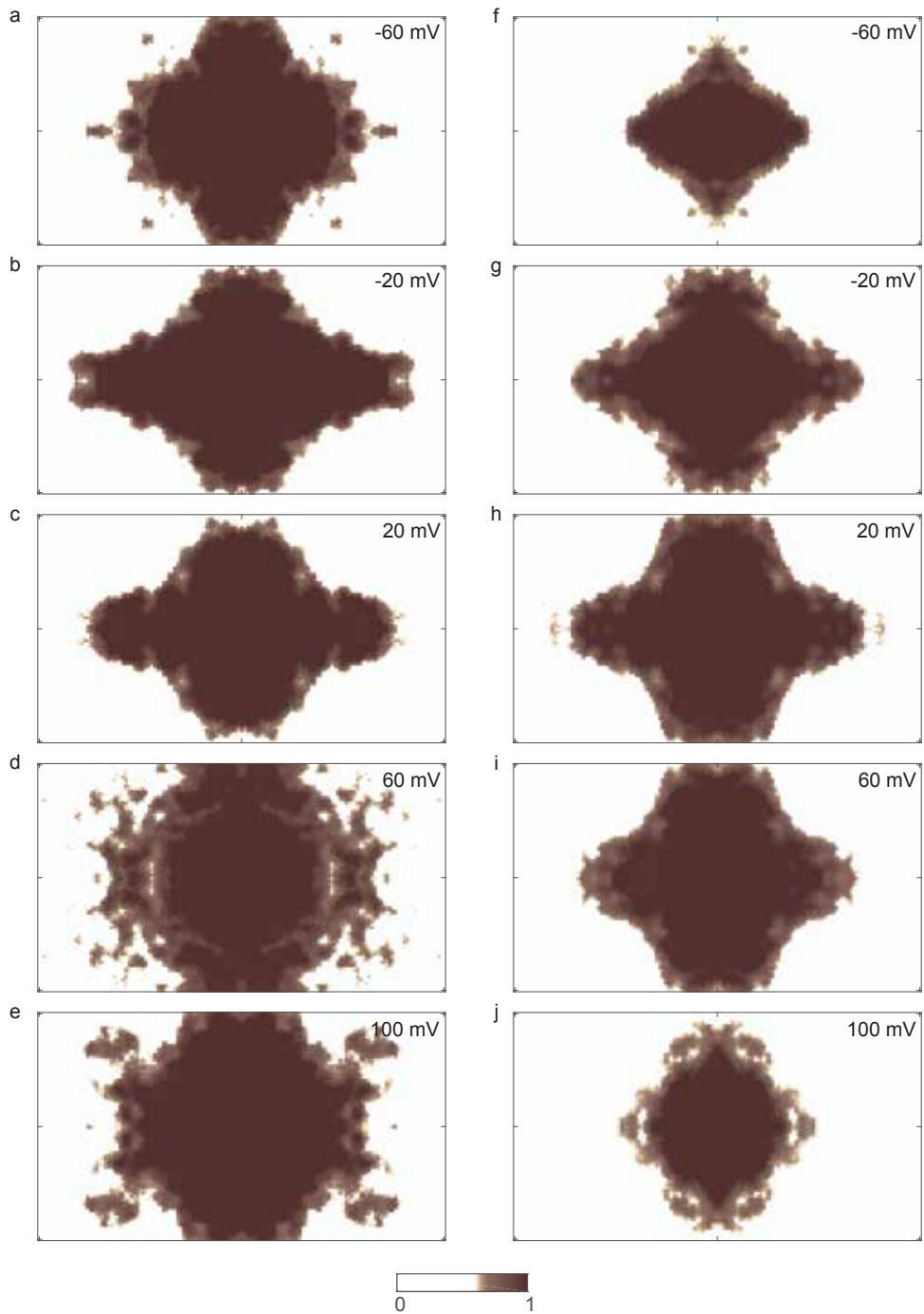

**Figure S10** - QPI, symmetrized and cropped to the first Brillouin zone, to the range $q_x = (-0.5\pi/a, 0.5\pi/a)$ and $q_y = (-0.5\pi/b, 0.5\pi/b)$. The data are taken at low temperature, in the Td phase, and normalized at V = 300 mV, I = 200 pA. (a-e) show the original data, and (f-j) show the data multiplied by the derivative of the Fermi-dirac distribution for 300K to simulate room temperature energy broadening. While fewer details are apparent in the QPI with energy broadening, the distinctive wing features are still strongly apparent.

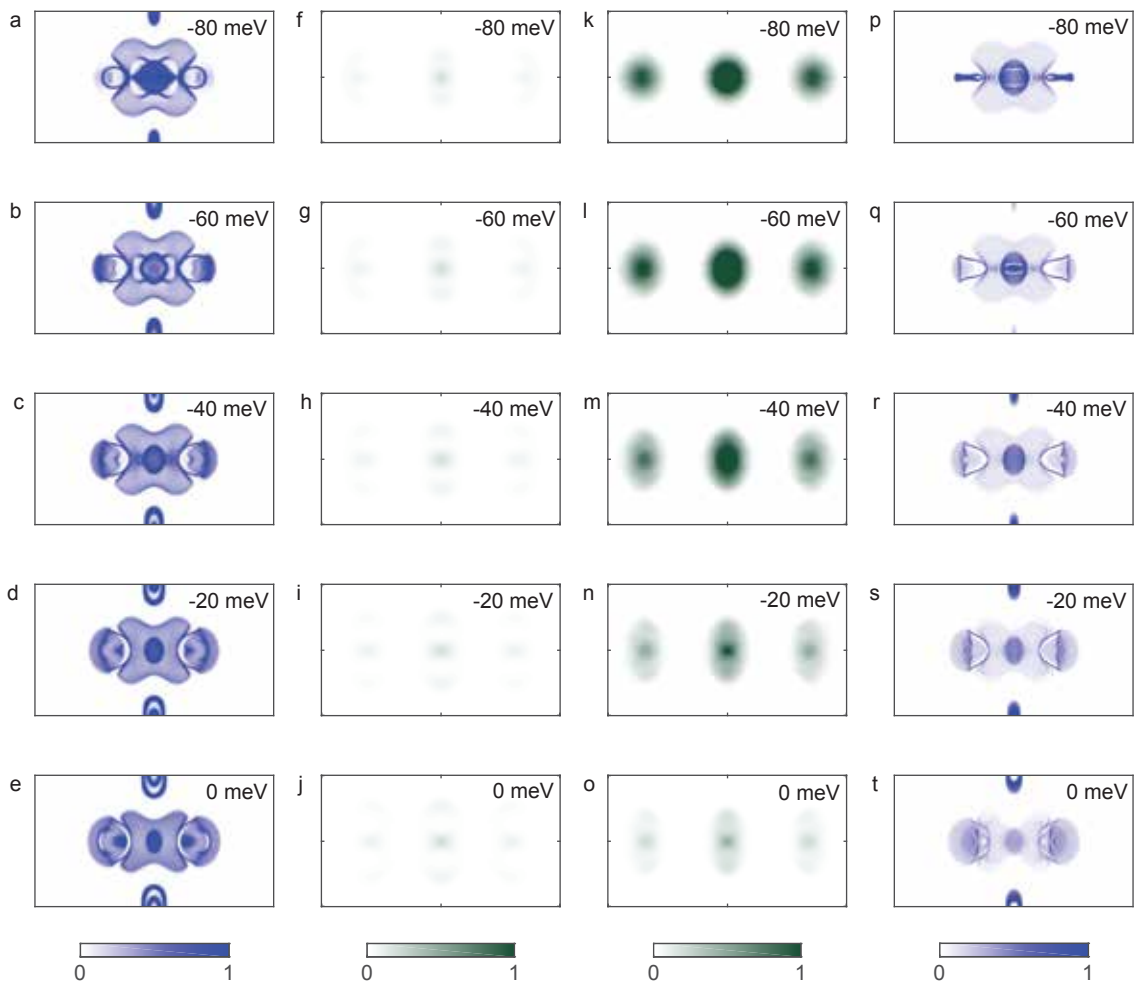

**Figure S11** - **(a-e)** Fermi surface for various energies of tight binding band structure calculated for the monoclinic phase, at room temperature. **(f-j)** QPI calculated from the tight binding band structure for the monoclinic phase. **(k-o)** QPI calculated from the tight binding band structure for the orthorhombic phase. **(p-t)** Fermi surface for various energies of tight binding band structure calculated for the orthorhombic phase, at low temperature. Each plot shown in this figure is cropped to the first Brillouin Zone, to the range $k_x$ or $q_x$ = (-0.5π/a, 0.5π/a) and $k_y$ or $q_y$ = (-0.5π/b, 0.5π/b) for the Fermi surfaces and QPI respectively. While both phases show QPI with similar features, the QPI in the room temperature phase is much less intense.

**Supplementary References:**